\documentclass[conference, anonymous = True]{IEEEtran}
\IEEEoverridecommandlockouts
\usepackage{cite}
\usepackage{amsmath,amssymb,amsfonts}
\usepackage{algorithmic}
\usepackage{graphicx}
\usepackage{textcomp}
\usepackage{xcolor}

\usepackage{subcaption}
\usepackage{booktabs}
\usepackage{makecell}

\makeatletter

\usepackage{fancyhdr}   

\fancypagestyle{firstpage}{     
  \fancyhf{}
  \chead{\textcolor{gray}{This article has been accepted for publication in the proceedings of the \\ The 23rd International Conference on Pervasive Computing and Communications (PerCom 2025) - Workshop PerFail.}}
  \fancyfoot[C]{\small{\textcolor{gray}{~\copyright~ 2025 IEEE.  Personal use of this material is permitted.  Permission from IEEE must be obtained for all other uses, in any current or future media, including reprinting/republishing this material for advertising or promotional purposes, creating new collective works, for resale or redistribution to servers or lists, or reuse of any copyrighted component of this work in other works.}}}

}

\def\BibTeX{{\rm B\kern-.05em{\sc i\kern-.025em b}\kern-.08em
    T\kern-.1667em\lower.7ex\hbox{E}\kern-.125emX}}
\begin{document}

\title{Collaborative Human Activity Recognition with Passive Inter-Body Electrostatic Field
}

\author{\IEEEauthorblockN{Sizhen Bian}
\IEEEauthorblockA{\textit{DFKI} \\
Kaiserslautern, Germany \\
sizhen.bian@dfki.de}
\and
\IEEEauthorblockN{Vitor Fortes Rey}
\IEEEauthorblockA{\textit{DFKI} \\
Kaiserslautern, Germany \\
vitor.fortes\_rey@dfki.de}
\and
\IEEEauthorblockN{Siyu Yuan}
\IEEEauthorblockA{\textit{RPTU} \\
Kaiserslautern, Germany \\
syuan@rptu.de}
\and
\IEEEauthorblockN{Paul Lukowicz}
\IEEEauthorblockA{\textit{DFKI} \\
Kaiserslautern, Germany \\
paul.lukowicz@dfki.de}
}

\maketitle

\begin{abstract}

The passive body-area electrostatic field has recently been aspiringly explored for wearable motion sensing, harnessing its two thrilling characteristics: full-body motion sensitivity and environmental sensitivity, which potentially empowers human activity recognition both independently and jointly from a single sensing front-end and theoretically brings significant competition against traditional inertial sensor that is incapable in environmental variations sensing. While most works focus on exploring the electrostatic field of a single body as the target, this work, for the first time, quantitatively evaluates the mutual effect of inter-body electrostatic fields and its contribution to collaborative activity recognition. A wearable electrostatic field sensing front-end and wrist-worn prototypes are built, and a sixteen-hour, manually annotated dataset is collected, involving an experiment of manipulating objects both independently and collaboratively. A regression model is finally used to recognize the collaborative activities among users. Despite the theoretical advantages of the body electrostatic field, the recognition of both single and collaborative activities shows unanticipated less-competitive recognition performance compared with the accelerometer. However, It is worth mentioning that this novel sensing modality improves the recognition F-score of user collaboration by 16\% in the fusion result of the two wearable motion sensing modalities, demonstrating the potential of bringing body electrostatic field as a complementary power-efficient signal for collaborative activity tracking using wearables. 

\end{abstract}


\section{Introduction}

\thispagestyle{firstpage} 

Wearable sensors are increasingly captivating researchers for a better understanding of human behaviours and correspondingly provide assistive services\cite{yan2022emoglass, van2020presley, bian2020wearable}.  Recent advances in wearable sensors have enabled a broader range of functional support in the field like sport\cite{tanaka2023full}, healthcare\cite{bian2019passive, kuosmanen2022does}, clinic\cite{parker2018interplay}, gaming\cite{lai2019fitbird}, interaction \cite{bian2021capacitive, azim2022over}, etc. 
Among them, wearable motion sensing has long been dominated by inertial sensors like accelerometers and gyroscopes, which play almost the unique role for motion sensing in current wearable devices \cite{bian2022state, wang2021nod}. However, one limitation of an inertial sensor is that it primarily perceives the motion of the body part to which it is attached and is limited in surrounding variation sensing (incapable) and cross-body part sensing (weak). In cross-body part sensing, the motion message of the targeted body part gradually gets lost when reaching the sensor-attached body part, e.g., the inaccuracy of step counting by a wrist-worn inertial sensor \cite{toth2019effects, bian2024earable}. 

To this end, a novel wearable motion sensing technique has emerged to supply complementary sensing ability to the inertial sensor, the body-area electrostatic field, also named human body capacitance, inferring body motions by observing the body surface charge flow pattern, which strongly relates to the body motion pattern \cite{cohn2012ultra}. 
Unlike other physiological features like electrocardiograph, the body-area electrostatic field is a feature that interacts with the surroundings, especially the ground. 
In substance, the body-area electrostatic field describes the charge distribution difference between the body surface and the surroundings. When inferring motion information from this feature, the field shows two thrilling characteristics: full-body movement detection, meaning the deployment of the sensing unit on the targeted body part is not mandatory. For example, a wrist-worn body electrostatic field sensing unit can track leg-based exercises even when the wrist is absolutely static \cite{bian2022using}; and surrounding motion detection, where the field sensing unit is sensitive to the body's interaction with its surroundings, including both contacting and being in the immediate proximity of people and objects \cite{bian2024body}.
Despite its subtlety, researchers have tried to make use of this intrinsic property for potential well-being services. Such explorations go back to early work on capacitive intra-body personal area networks \cite{zimmerman1996personal}, followed by proximity sensing \cite{zimmerman1995applying}, communication \cite{cohn2012humantenna}, human-machine interaction \cite{suh2023worker} and motion detection \cite{cohn2012ultra, bian2019wrist, bian2022exploring}. 
Nevertheless, most existing works exploring wearable body electrostatic field sensing only focus on single-body activity perception or body-machine interaction, and body-to-body cooperation was not attentively explored. 
When multiple bodies collaboratively take action and are in the near range of each other, their body-area electrostatic field will be mutually affected; thus, extra collaborative contextual knowledge could be potentially deduced besides the traditional wearable motion sensing, leveraging the feature of full-body and surrounding sensitivity of the body electrostatic field. Based on this expectation, we moved one step further from the existing works of recognition of physical actions performed by an individual to physical actions collaboratively performed by group users \cite{gruenerbl2017detecting} and tried to recognize physical activities that are a challenge for the classical wearable motion sensor.

\section{Related Works}

Existing works in collaborative activity recognition were mainly based on camera surveillance videos, which focused on spatiotemporal relations among the human bodies in the scene and deduced collaborative activities through tracking multi-agent spots\cite{zhang2008hierarchical,li2009learning}. Video-based group activity recognition suffers from its high computational cost and other ethical issues. This work focuses on wearable sensor-based solutions and aims to facilitate the detection of joint physical manipulation of objects that are more difficult to detect visually.
Concerning the sensor-based analysis of multi-agents activities, Wilson et al. \cite{wilson2005simultaneous} offered a combination of motion detectors, break-beam sensors, pressure mats, and contact switches to detect agents' proximity and touch actions. An infrared sensor-based scalable network \cite{wren2006toward} was also proposed for perceiving people's activities in an intelligent building and got better than 90\% recognition performance for low-level activities. Wang et al.\cite{wang2009sensor} developed a multi-modal wearable sensing platform and presented a theoretical framework to recognize both single-user and multi-user activities, in which temperature, humidity, light, audio, RFID, motion sensors were collected to contribute to the recognition. 
Gordon et al. \cite{gordon2013towards} explored a node with motion sensors, which can be attached to a mug, by deploying several nodes on mugs; the collaborative group activity was extracted at mobile devices' side. 
Those multi-modal sensor-based multi-agent activity recognition platforms were heavy to deploy, and they usually focused on detecting whether the involved agents were gathering or dispersing as an indication of group activities. A high-level activity like agent collaboration is a more challenging topic and exists in plenty of practical scenarios, especially in manufacturing \cite{ward2017detecting, stiefmeier2008wearable, moutinho2023deep}. 

\begin{figure}[]
\centering
\includegraphics[width=0.25\columnwidth,height=3cm]{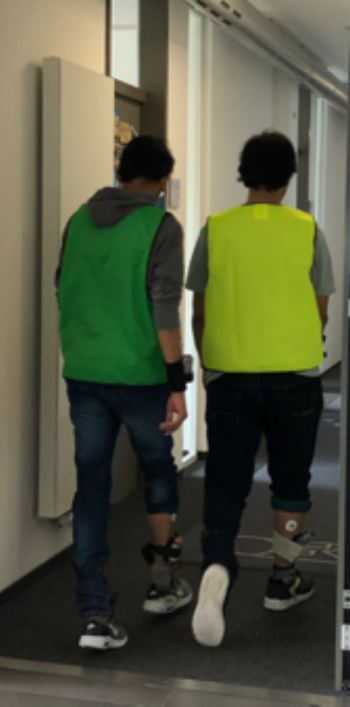}
\includegraphics[width=0.25\columnwidth,height=3cm]{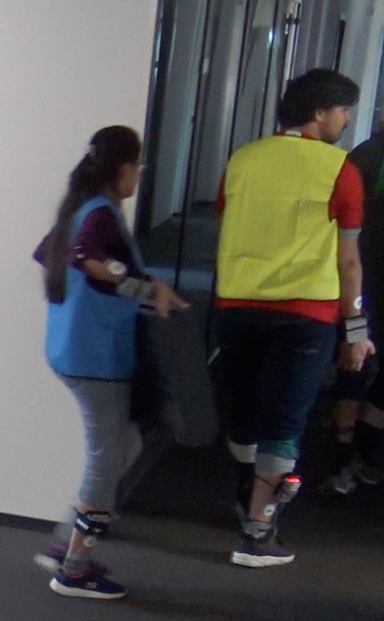}
\includegraphics[width=0.25\columnwidth,height=3cm]{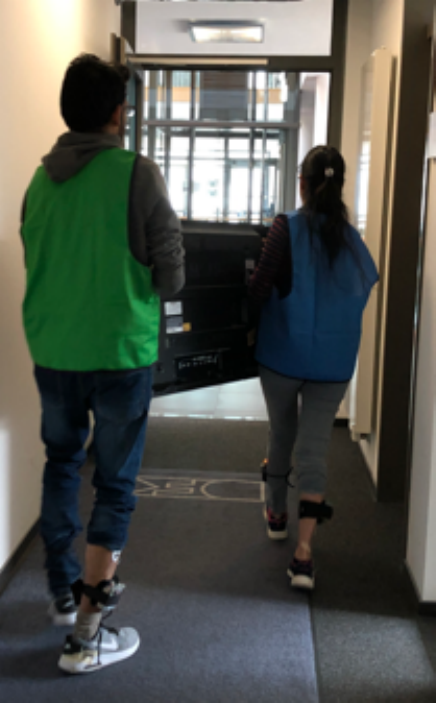}
\caption{Example of the type of recognition tasks that our work partly targets: distinguishing group agents among normal walking next to each other, walking next to each other and operating an object alone, and walking next to each other while jointly operating an object. Co-location is not a solid separator, and camera filming is often limited by line of sight.}
\label{capacitance_model}
\label{fig:problem}
\end{figure}

Benefiting from the intrinsic property of the body-area electrostatic field and its low power and small form factor sensing unit, this work explores this novel sensing modality for recognizing the typical scenario of physical works related to manipulating objects and physical collaboration between users. The problem we addressed is partly illustrated in Figure \ref{fig:problem}. Previous work has investigated using inertial sensors and co-location information for such recognition tasks \cite{roggen2010collecting}. The problem is that the motion and posture differences involved in, for example, handling an object jointly vs. handling an object alone are subtle and often overshadowed by differences resulting from the specific object being handled and inter-person variations. Furthermore, people just walking next to each other may be closer in terms of location than people jointly manipulating a large object, so co-location is also not a conclusive indication of collaboration.

Overall, we present the following contributions in this paper:
\begin{enumerate}
\item We designed an ultra-low-power, low-cost, small form factor, wearable body-area electrostatic field sensing front end and the corresponding prototype aiming to evaluate the value of body-area electrostatic field sensing for collaborative group activity recognition.
\item The study of a collaborative experiment, TV-Wall assembling and dissembling, indicated an unanticipated less-competitive recognition performance of electrostatic sensing compared with the accelerometer and, meanwhile, an impressive improvement in user collaboration recognition by 16\% to the accelerometer-based solution. Such results have enlightened us with a further in-depth exploration of this novel motion sensing modality, mainly on the hardware side.

\end{enumerate}

\begin{figure}
\begin{minipage}[t]{0.45\linewidth}
\centering
\includegraphics[width=0.8\textwidth,height=2.0cm]{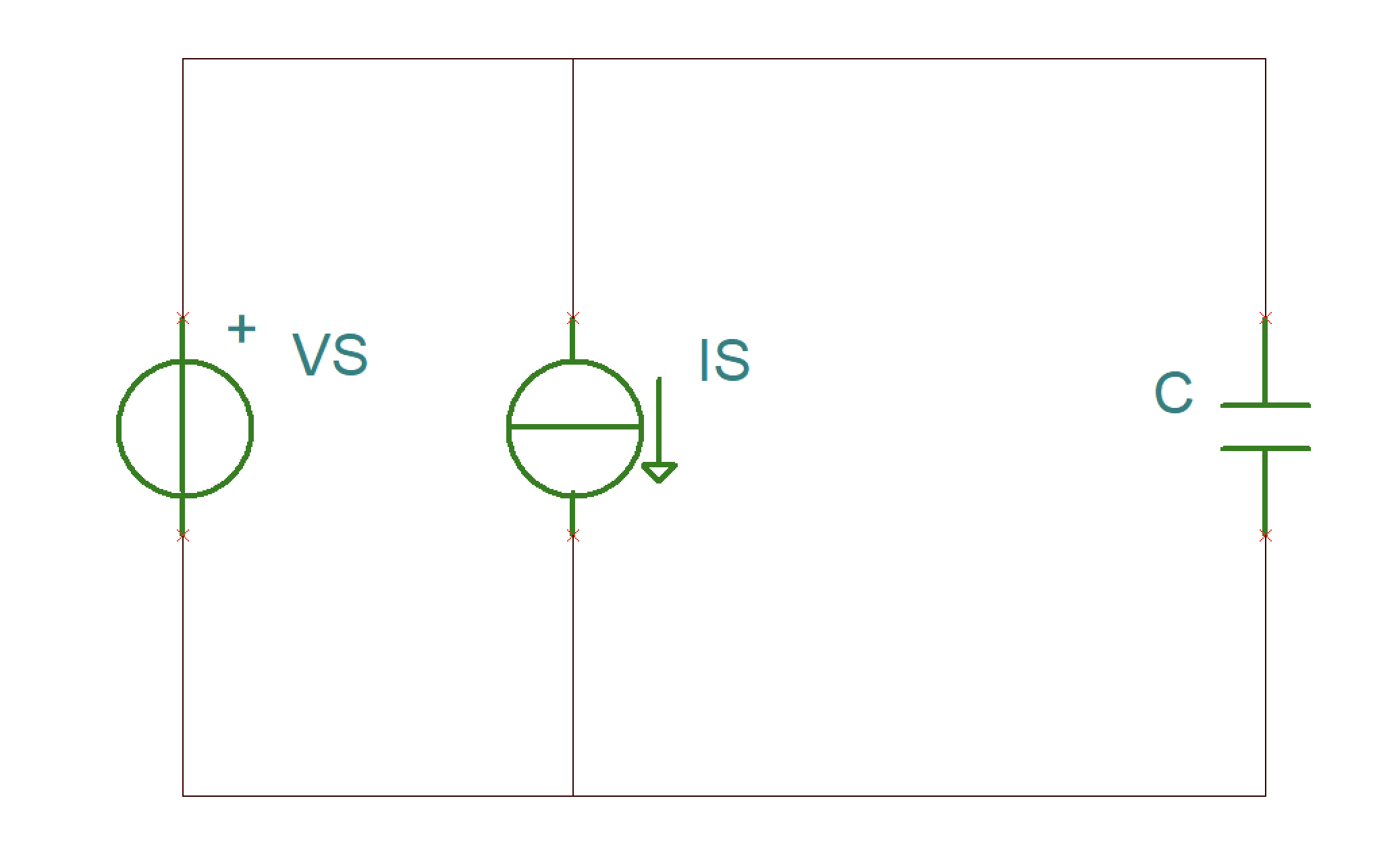}
\caption{Basic structure of a body capacitance sensing method}
\label{Sensor}
\end{minipage}
\quad
\begin{minipage}[t]{0.45\linewidth}
\centering
\includegraphics[width=0.95\textwidth,height=2.0cm]{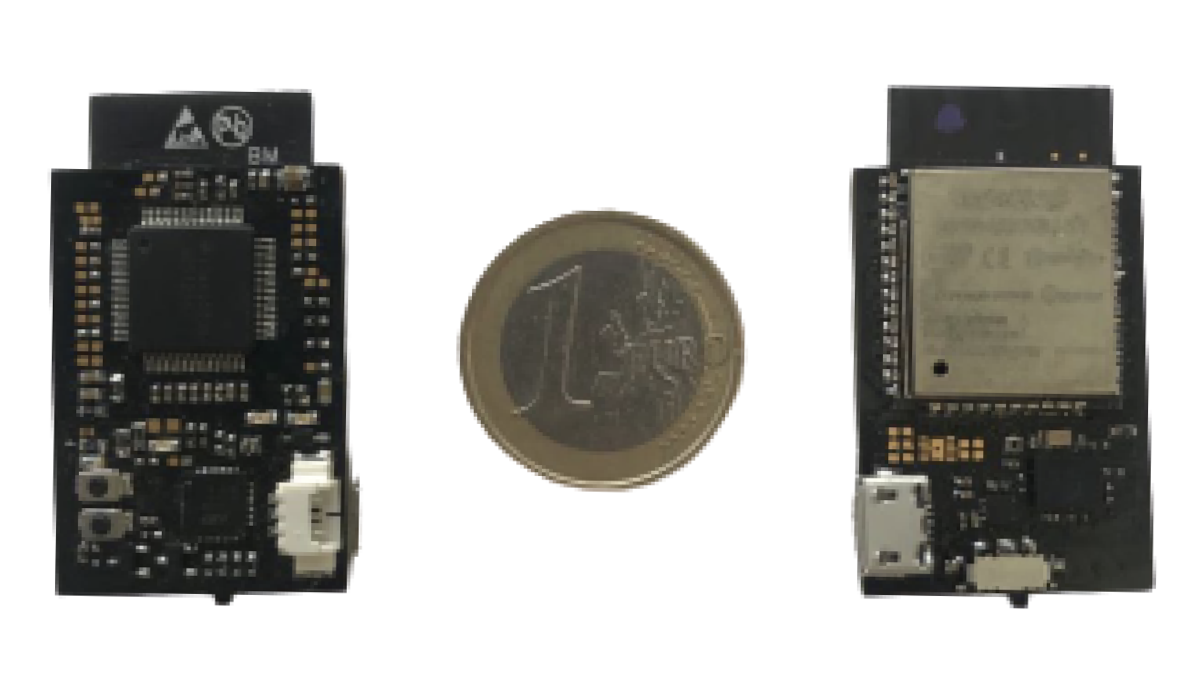}
\caption{A wearable prototype for body motion sensing}
\label{Prototype}
\end{minipage}
\end{figure}

\section{Sensing Prototype}
Instead of monitoring the body capacitance directly, we observe the body's potential as an indication. Variation in body capacitance will cause a potential instantaneous change on the body surface. Figure \ref{Sensor} shows the basic principle of our sensing front-end design. The voltage source maintains the potential level on the body surface; the current source supplies electrons to $C$. Once $C$ varies, a potential variation will occur immediately. After a while, the potential level returns to $VS$ with the complement of the charges from $IS$. The whole mechanics is a series of charging and discharging processes. Figure \ref{Prototype} shows our prototype for body motion tracking, comprising of a battery charging module, an ESP32 processing unit with WIFI and BLE integrated, a 24-bit high-resolution ADC ads1298, an inertial sensor, and the electrostatic sensing front end. The sensing front-end, prototype schematic, and all other materials (annotated dataset, applied models, etc.) in this work is publicly released for further study.

\section{Evaluation}

\begin{figure*}[t!]
    \centering
    \begin{subfigure}[t]{0.23\textwidth}
        \centering
        \includegraphics[width=0.8\textwidth,height=2.5cm]{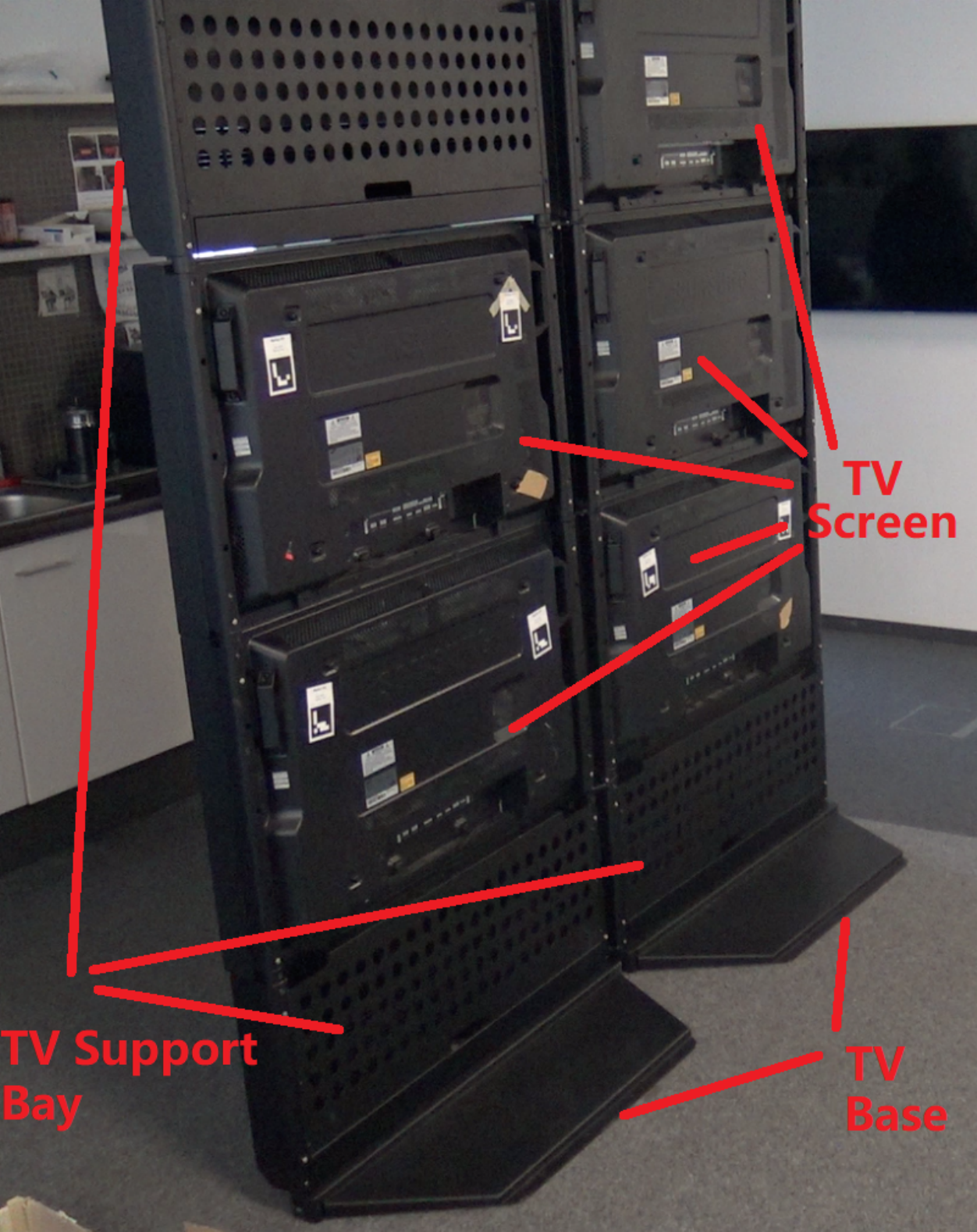}
        \caption{2.44m High TV-Wall}
        \label{TV_Wall}
    \end{subfigure}%
    ~ 
    \begin{subfigure}[t]{0.23\textwidth}
        \centering
        \includegraphics[width=0.8\textwidth,height=2.5cm]{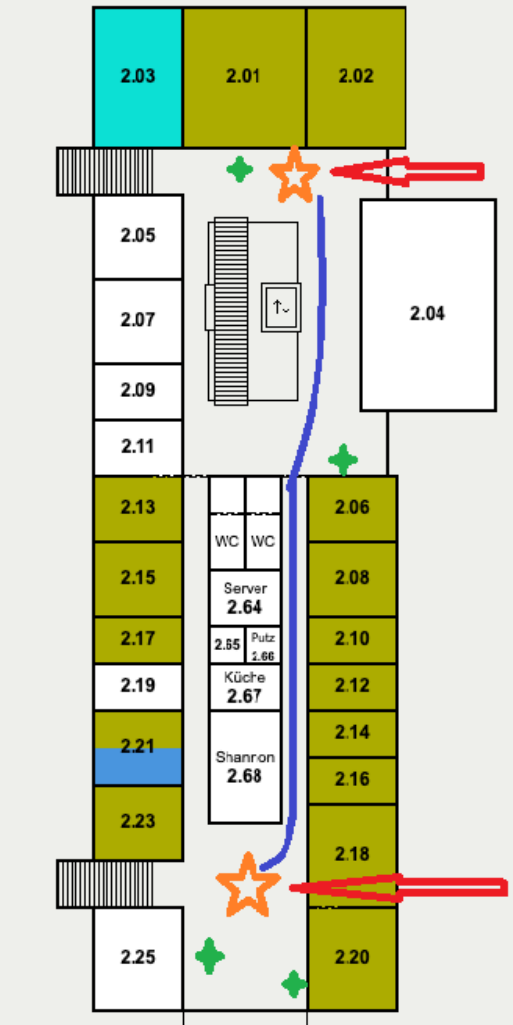}
        \caption{Collaborative activity's map}
        \label{Building}
    \end{subfigure}
    ~ 
    \begin{subfigure}[t]{0.23\textwidth}
        \centering
        \includegraphics[width=1.0\textwidth,height=2.5cm]{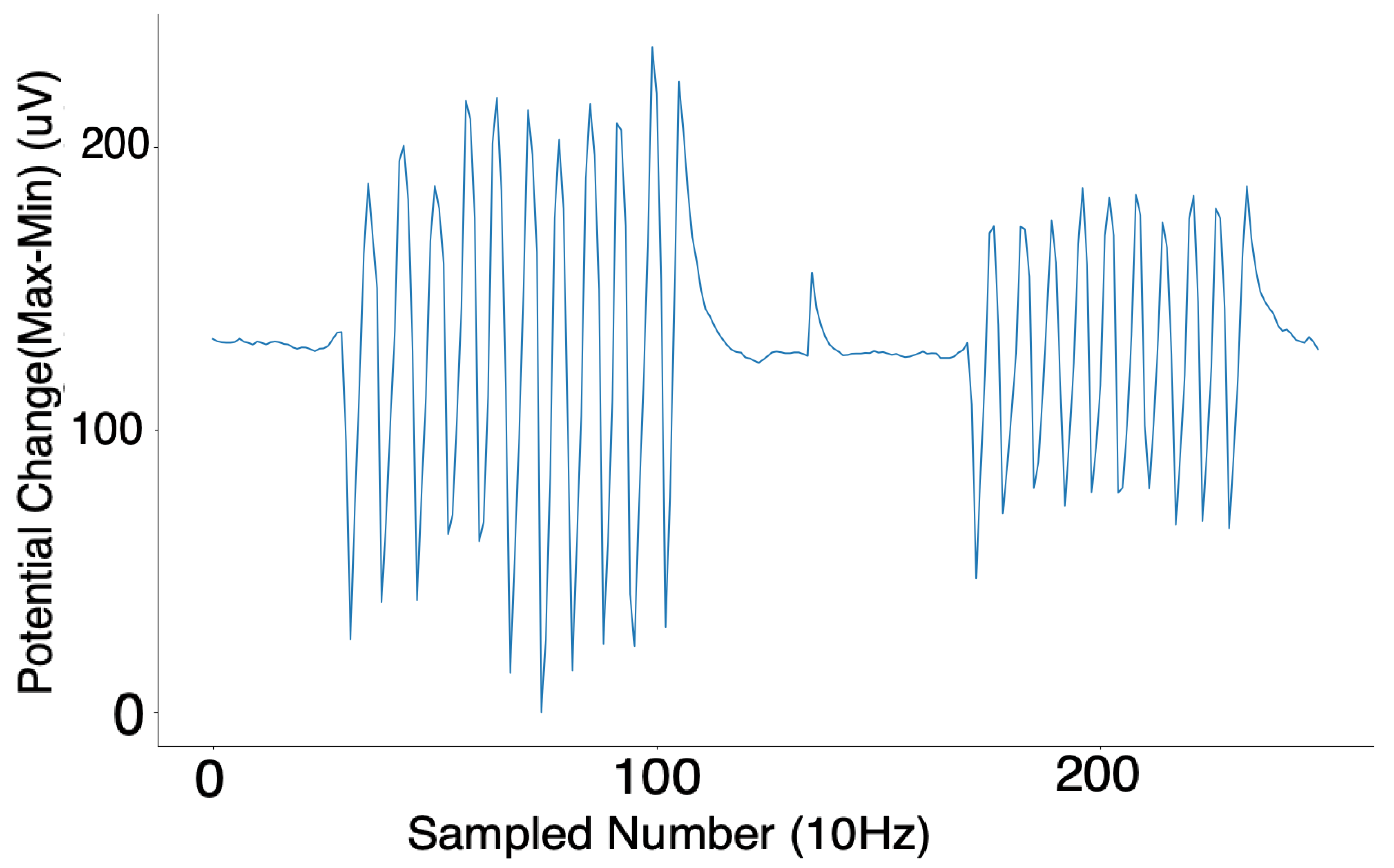}
        \caption{Walking alone and Joint Walking}
        \label{Walk_Defference}
    \end{subfigure}%
    ~ 
    \begin{subfigure}[t]{0.23\textwidth}
        \centering
        \includegraphics[width=0.8\textwidth,height=2.5cm]{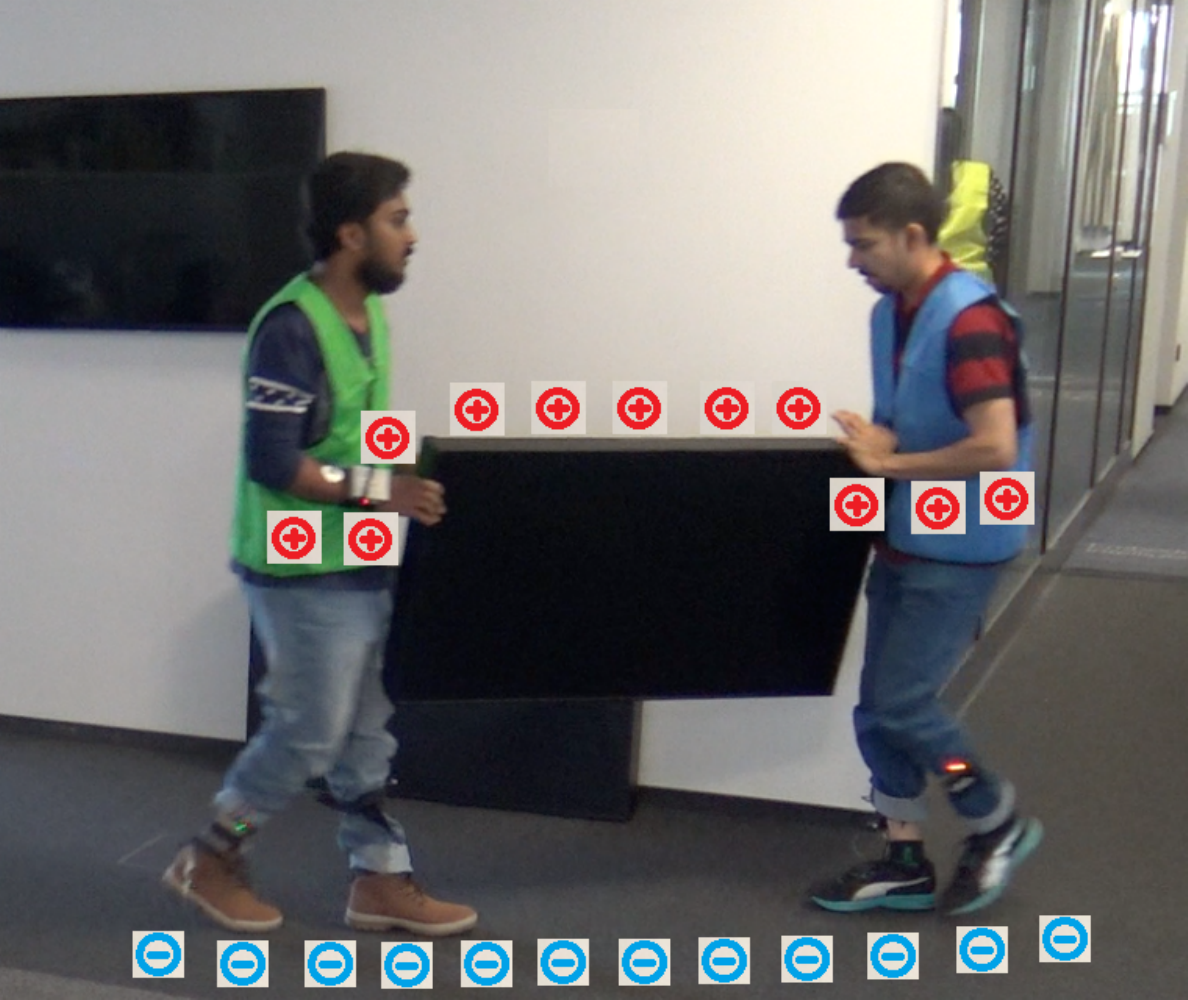}
        \caption{Electrically coupling between multi-agents}
        \label{Taking_TV}
    \end{subfigure}
    ~ 
    \\
    \quad
    \\
    \caption{Collaborative Work and the Sensing Modality}
    \label{HBC_Sensing}
\end{figure*}

To evaluate body electrostatic field sensing in the wearable form for group activity recognition, we designed a collaborative physical work including both independent and joint activities of each worker, assembling and disassembling a TV-Wall. Twelve participants(ten male and two female) were divided into four groups. Each group moved the giant TV screens from the storeroom to a task operating spot, assembled and disassembled a TV wall, and carried them back. They performed this physical task 4 times in 4 days, and the operation lasted around one hour per time. As Figure \ref{TV_Wall} presents, the 2.44 m TV-Wall comprises 3 screen support bays weighing 10.3 kg each, 2 TV bases weighing 22.1 kg each, and 5 TV screens weighing 23.2 kg each. The lighter one can be carried by a single participant. For the others, two participants are needed to take and carry jointly. Figure \ref{Building} depicts the map where the activities are performed. The orange signs and red arrows indicate those heavy metals' original and operational location. Participants must carry them from the top orange sign spot to the lower orange sign spot following the blue line. The route is around 36 meters. We used four cameras (placed at the green spots in the figure) to record the whole working process for annotation; every participant knew and agreed to the presence of cameras.   
In the experiment, participants wore one prototype on their preferred wrist and did the task naturally without instruction. Finally, we got 39 sessions of valid data altogether; each contains around one hour's motion signals from the wrist-worn electrostatic field sensing unit and accelerometer. An extra accelerometer was placed on the calf to evaluate the contribution of electrostatic field sensing to a different signal source, aiming to increase the robustness of the potential positive conclusion.
Figure \ref{Walk_Defference} briefly depicts the benefits of utilizing body electrostatic field-induced signal for observing joint activity like simple walking. The left signal was perceived when person A (wearing the sensing prototype) was walking normally, and then he shook hands with person B (the moment the middle impulse occurs). While strongly electrically coupled, person A made another walk around person B. Through coupling, the body capacitance will be enlarged, resulting in the decrease of surface potential change. When multiple people are electrically well coupled or connected, charges on both bodies can flow until a balanced level is reached(Figure \ref{Taking_TV}). This feature contributes to recognizing if agents are working collaboratively or separately.

\subsection{Activity classes}

\begin{table}[htb]
\caption{Type of Activities}\label{ActivityType}
\centering
  \begin{tabular}{|p{0.6cm}|p{1.8cm}|p{5.2cm}|}
    \toprule
    \small\textit{ID}&\small\textit{Activities}&\small\textit{Comments}\\
    \midrule
    A1&Start and Stop & 10 on-site steps \\
    A2&Doing nothing & stand still without any movement\\
    A3&Normal Walk & normal walking without carrying anything\\
    A4&Carry alone & walking and carrying the 10.3kg metal pieces\\
    A5&Carry jointly & walking and carrying the over 20kg metal pieces with another person\\
    A6&Lift & touch and lift the metal pieces from the box,ground and TV Wall\\
    A7&Drop & drop the metal pieces into the box, on ground and to the TV Wall\\
    A8&Turn screw & turn the screws with an electric screwdriver \\
    A9&No definition & none of above, e.g. drinking\\
    A10&Out of camera & walked out of camera's field of view \\
    \bottomrule
\end{tabular}
\end{table}

Each session was divided into ten primitives A1-A10, as Table  \ref{ActivityType} lists. A1 aims to synchronize sensor data. A2 mainly occurred when the participants took a rest. A3, A4, and A5 are the most relative primitives to evaluate this novel sensing modality. Carrying jointly means the participants are electrically coupled or connected, causing the charge redistribution on both bodies. A4 is an independent activity with a heavy load at hand, which differs from A3 while the load enlarges the conductive plate at the body side. Lift and Drop are primitives that can happen independently or jointly since those two activities have only a limited motion distance between arms and ground, so there is not too much difference in the electrostatic field signal whether they were performed individually or jointly. 
A8 was performed with an electric screwdriver, which only needed a fingertip movement, so this activity did not generate a useful signal.
Four cameras were deployed to record the whole task process. After synchronizing the videos, we labeled the data for each participant manually. 

\begin{figure*}
\centering
\includegraphics[width=1.0\textwidth,height= 6.0cm]{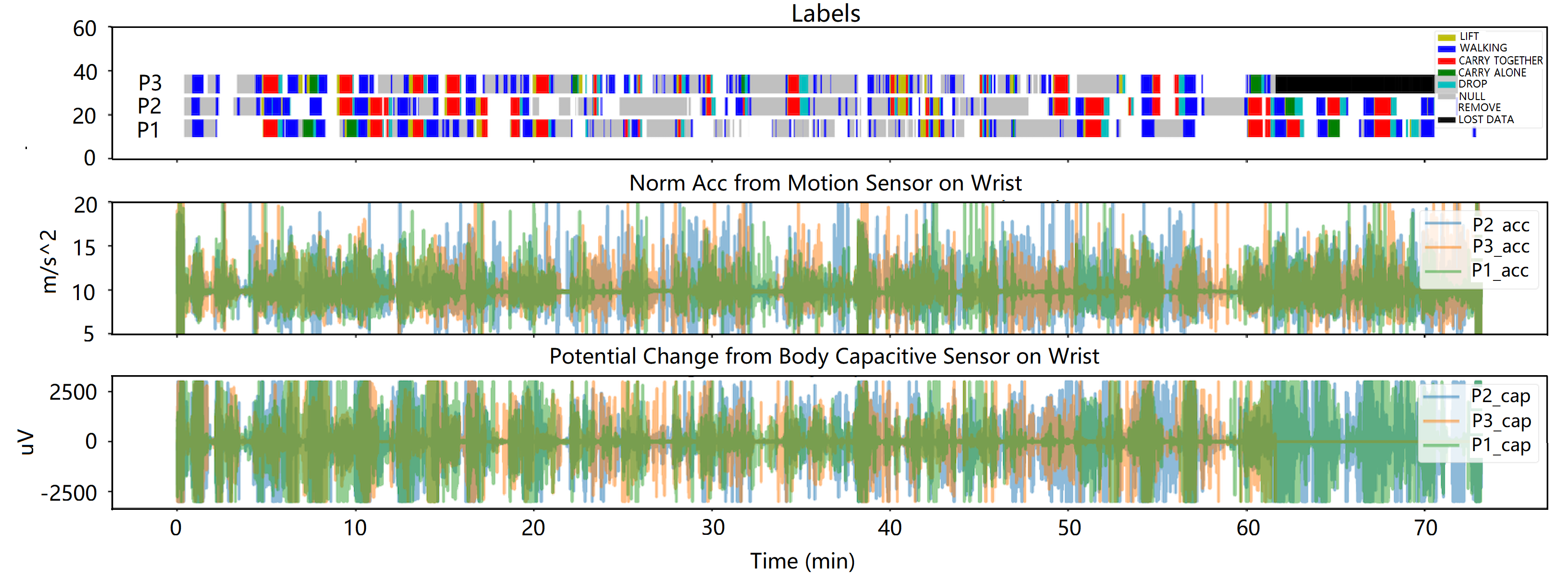}
\caption{Example Session of Labeling and Preprocessed Signal}
\label{Labelling}
\end{figure*}

\subsection{Classification Exploration}
We generate instances by applying a $5$ second sliding window with a $1$ second step. Labels for each window are determined using majority voting inside it.
We use the norm of the accelerometer data and the changes in potential value instead of their raw values. The absolute maximum change from the electrostatic field sensing front end allowed is 3mV, and any more massive change is considered an outlier and replaced by either 3mV or -3mV, according to its direction. 


\noindent We performed activity recognition in two modalities:
\begin{itemize}
    \item Receiving test data from a single user and predicting the activities of A3, A4, A5, A6, A7. Activities of A2, A8 are moved into a new class named null class. 
    \item Receive test data from both users and predicting the collaborative primitives of A5, A6(together), A7(together), with A2, A3, A4, A6(alone), A7(alone), A8 being considered into the null class.
\end{itemize}

There are three cases in which the generated sliding windows were discarded. The first case is the activity of A9 and A10, which is beyond our research interest(A9, A1), and the ground truth is impossible to annotate(A10). The second case is where labeling information is missing, no related activities were performed or where the participants were out of the camera, those intervals are marked as white. The last case is data loss caused by some occasional hardware problem, data was failed being written into SD card. This can be seen in the upper part of Figure \ref{Labelling} where missing data is marked as black. The middle and lower sections of Figure \ref{Labelling} show the pre-processed data of the accelerometer and electrostatic field signal.


For classifying the windows, we trained different machine learning models, and the logistic regression model using one versus all gave the best result. Since the data set is imbalanced, containing more null class instances than other activity types, every training instance is weighted based on the labels present inside the window. 
Classifier predictions are smoothed by deciding the label for each window using soft voting where the current window, $3$ windows forward and backward vote, 
which helps to smooth predictions.





\begin{figure*}[t!]
    \centering
    \begin{subfigure}[t]{0.45\textwidth}
        \centering
        \includegraphics[width=1.0\textwidth,height=3.7cm]{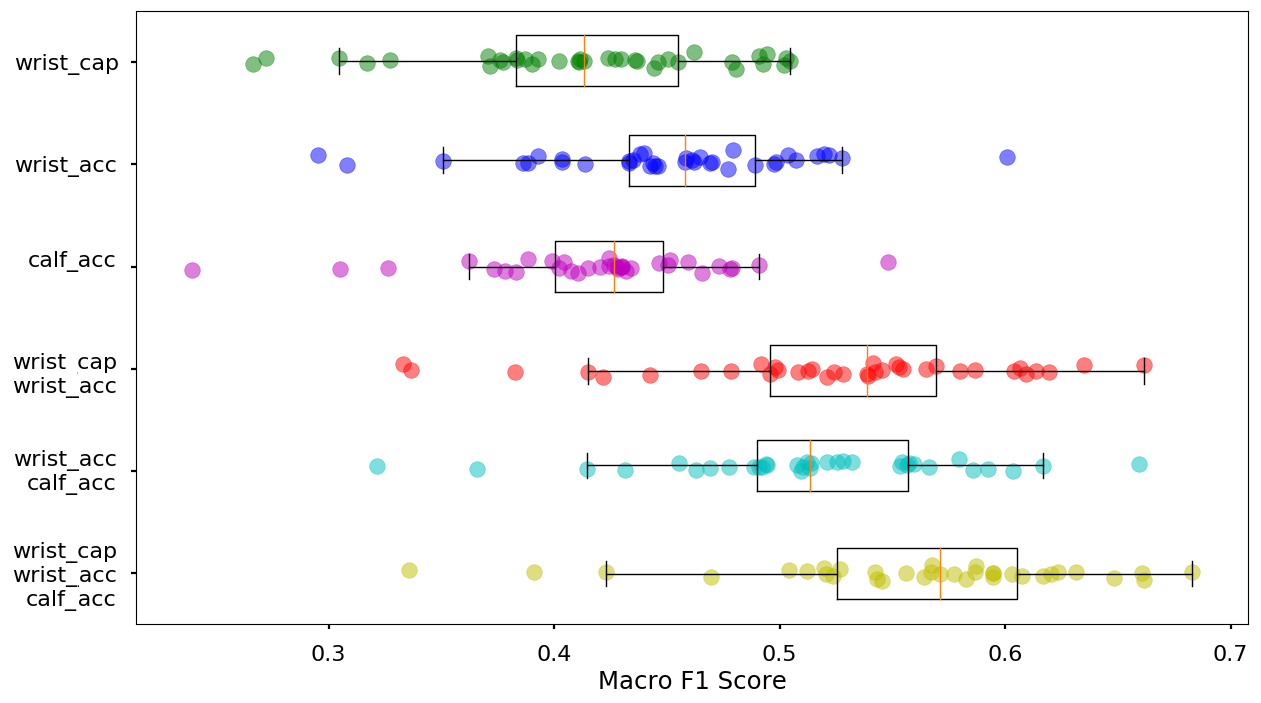}
        \caption{Receiving data from single user}
        \label{Data_Single_HARD}
    \end{subfigure}
    ~
    \begin{subfigure}[t]{0.45\textwidth}
        \centering
        \includegraphics[width=1.0\textwidth,height=3.7cm]{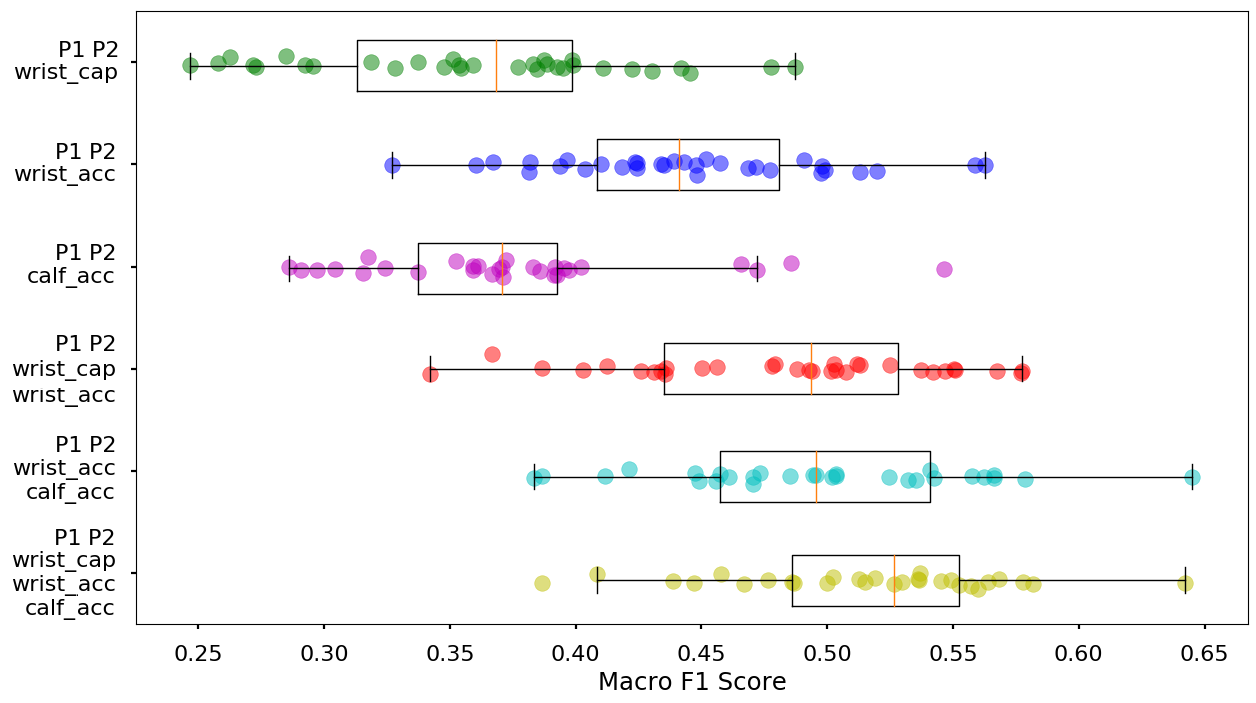}
        \caption{Receiving data from both users pairwise}
        \label{Data_Both_HARD}
    \end{subfigure}

    \caption{Macro F-Score with Lift/Drop as Separate Classes when using single Sensor and Sensor Fusion}
    \label{Sensorfusion_HARD}
\end{figure*}

\begin{figure*}[t!]
    \centering
    \begin{subfigure}[t]{0.45\textwidth}
        \centering
        \includegraphics[width=1.0\textwidth,height=3.7cm]{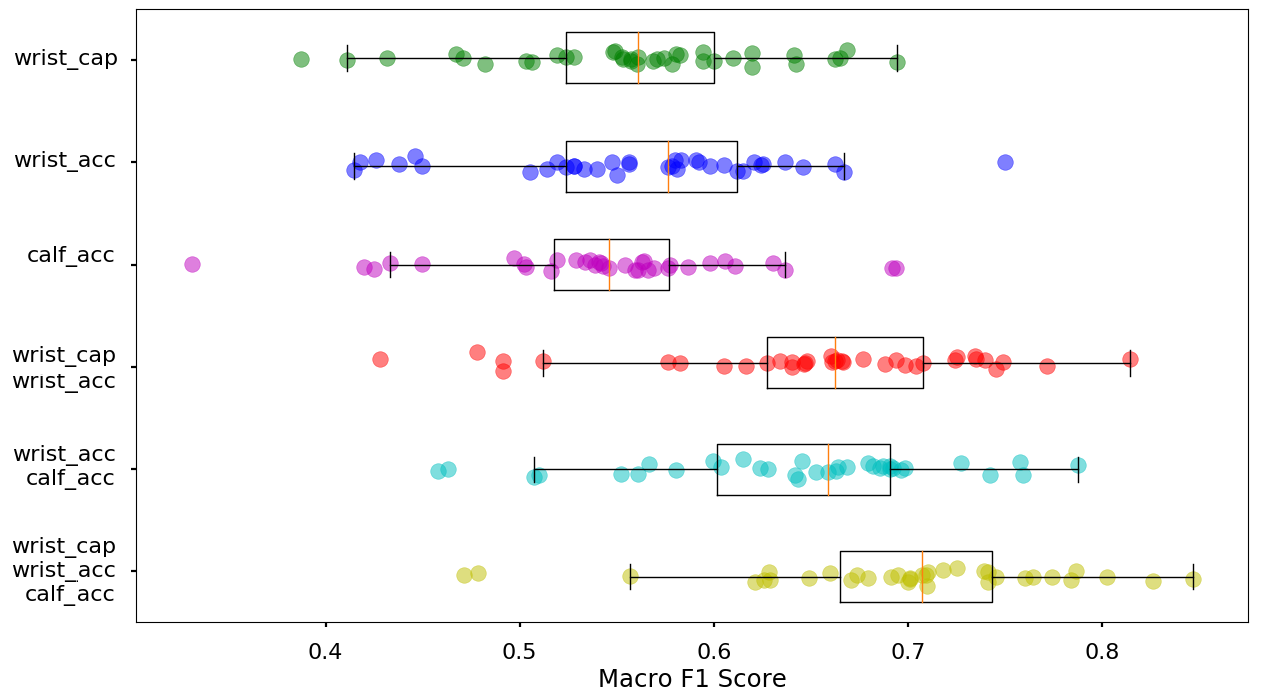}
        \caption{Receiving data from single user}
        \label{Data_Single}
    \end{subfigure}
    ~
    \begin{subfigure}[t]{0.45\textwidth}
        \centering
        \includegraphics[width=1.0\textwidth,height=3.7cm]{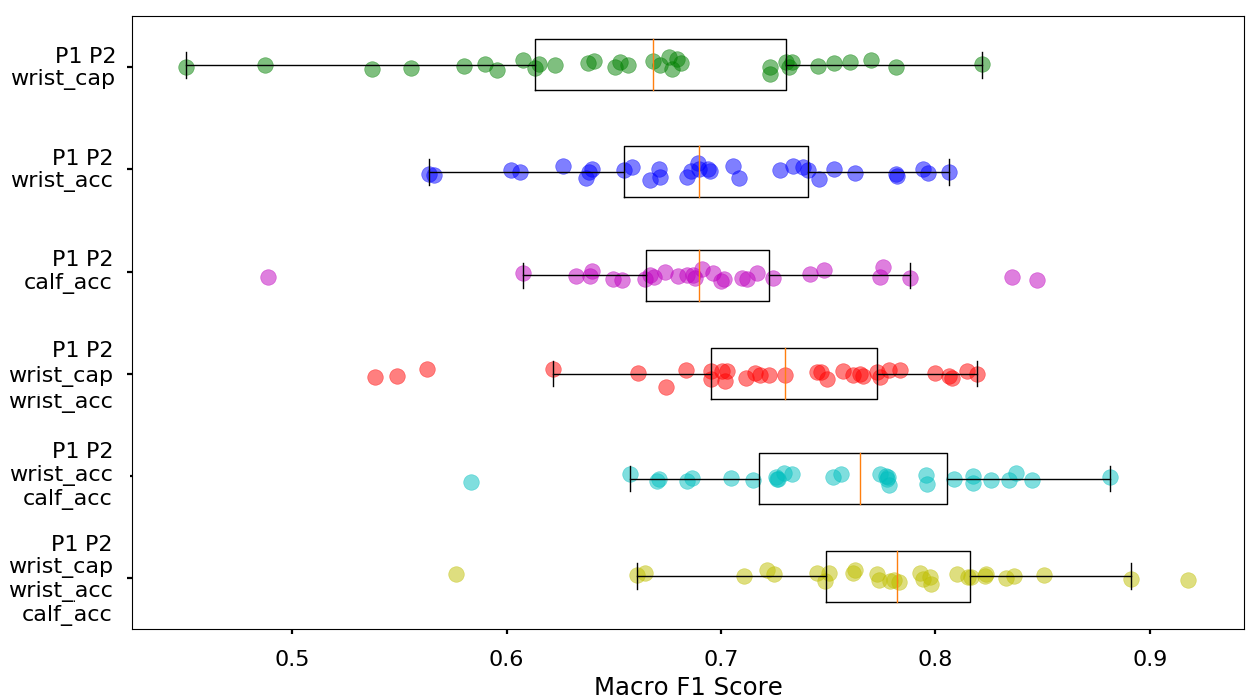}
        \caption{Receiving data from both users pairwise}
        \label{Data_Both}
    \end{subfigure}

    \caption{Macro F-Score with Lift/Drop into Null Class when using single Sensor and Sensor Fusion}
    \label{Sensorfusion}
\end{figure*}

\subsection{Classification Results}
Figure \ref{Sensorfusion_HARD} shows the classification result of different modalities with different deployments. To show that we can learn to recognize activities across groups, we employed a leave-one-group-out procedure where, for each fold, the test set contains all days of one group, while the training set contains all days for the remaining groups. 
When the test data was from a single participant, we got a combined macro F-score with 0.42, 0.46, and 0.43 when the data source was the electrostatic field sensing unit on the wrist, accelerometer on the wrist, and calf. This result is unanticipated as we expect an outperforming recognition performance of the body capacitance over the inertial sensor. The accelerometer is not theoretically able to recognize collaboration since it cannot perceive the actions of other nearby participants, unlike the body capacitance. In practice, the participant's action will vary slightly with the invasion of others. The classification ability of the calf accelerometer is located in the swing and stance phase, step and stride duration of the gait in the different action primitives. The accelerometer on the wrist also supplied the ability to perceive variant wrist motion models at a certain level. As for the electrostatic unit on the wrist, as we explained, $"$Carrying together$"$ means more than doubling of the body capacitance, $"$Carrying alone$"$ means enlarging it by the metal load in their hands, and $"$Walking alone$"$ keeps the human body capacitance as itself. All three sensor sources together contributed an F-score of 0.56. The right column of Figure \ref{Sensorfusion_HARD} shows the recognizing result when predicting only inter-group collaboration in non-scripted scenarios, which means we predicted here only primitives of $"$Carrying together$"$, $"$Drop together$"$, $"$Lift together$"$ and the others. The recognition was performed by receiving data from pairs of collaborators. Again, here, we predicted the recognition rate with different sensor sources and sensor fusion. When leaving $"$Drop$"$ and $"$Lift$"$ as independent classes, the capacitive sensor improved the combined F-score from 0.44 with a single wrist accelerometer to 0.47, and the calf accelerometer further assisted this value to 0.51.
We wonder how the recognition performs with only wrist-worn sensors, which are more comfortable to use in practical scenarios. Combining the two wrist sensors, we got an increased combined F-score of 0.53 receiving data from a single user and all the three primitives($"$Carrying together$"$, $"$Carrying together$"$ and $"$Walking alone$"$) got better recognition accuracy, meaning that the electrostatic sensing modality granted a raise of 0.07(15\%) in combined macro F-score for the single wrist accelerometer. When receiving data from both users, the recognition increase from the capacitive sensing was 7\%.

As we described in the activity classes, the primitives of Lift and Drop were time-short and motion-limitation actions(only the arms were stretching out and drawing back). The corresponding signal features were tough for our classifier to recognize. Thus, they were both frequently recognized as $"$Null$"$ class or mixed with each other. Another reason for this inaccuracy came from the manual labeling process. As in practice, the participants' motions were mixed; they dropped or lifted stuff while walking in many cases, and walking caused electrostatic variation signal easily overlap the drop or lift-caused signal. Our labeling result was time-in-series, the concurrency of the primitives was not considered. 
Since the most negative influence for recognizing was from the primitives of $"$Drop$"$ and $"$Lift$"$, and we were more interested in $"$Carrying together$"$, $"$Carrying alone$"$ or $"$Walking alone$"$, we moved this two classes into $"$Null$"$ state and got classification result of Figure \ref{Sensorfusion}. The classification result was highly improved compared with before. With all three data sources, we got a combined F-scores of 0.69 and 0.78 for each recognition modality separately. Each of them provided acceptable recognition of collaboration. When not considering the accelerometer signal from the calf, the wrist accelerometer gave 0.56 F-score when receiving data from single user, the wrist electrostatic unit improved this to 0.65(16\% increase).

In both data receiving approaches and every class configuration, the wrist-worn system(fusion of wrist accelerometer and wrist electrostatic sensor) benefits from the new sensing approach, which increases the single wrist accelerometer with 15\%, 7\%, 16\%, 4\% separately in the four situations. In conclusion, despite the unanticipated less-competitive recognition performance of electrostatic sensing compared with the accelerometer, the body-area electrostatic field sensing modality can still support the traditional motion sensor regarding recognizing collaborative activity. Limited by the page number, detailed classification numbers in the form of a confusion matrix will be provided on request.


\section{Discussion and Future Work}
This work described a wearable human body motion tracking prototype composed of two sensing modalities, the traditional accelerometer and a novel body-area electrostatic field sensing unit, to track individual and collaborative activities. The electrostatic sensing modality features itself with low cost and low power consumption and enjoys full-body and surrounding-sensitive advantages over the inertial sensor, thus potentially providing better collaborative activity recognition than the traditional inertial sensor. However, the evaluation result 
didn't meet our expectations. There are mainly two reasons for this less-competitive result: first, we overlooked the motion pattern differences of a single user when jointly and independently manipulating objects, and such differences could be perceived by the inertial sensor benefitting from its high sensitivity; second, we overestimated the robustness of the body-area electrostatic sensing as it is an interactive signal between surrounding and the human body and are sensitive to many influence factors. Despite this unanticipated result, the electrostatic field sensing is not worthless. By fusing the electrostatic sensing on the wrist and accelerometers on the wrist and calf, we achieved the best classification accuracy, e.g., 71\%, 64\%, 72\%, 88\% for walking alone, carrying the object alone, carrying the object jointly and the left null state respectively when receiving test data from single-user and 82\%, 91\% for carrying the object jointly and the other performed activities when receiving data from both users pairwise. Future work will be focused on the hardware side for a more sensitive and high signal-to-noise ratio front-end aiming to provide fundamental improvement of using the body-area electrostatic field for a novel and solid wearable motion sensing modality.



\bibliographystyle{IEEEtran}
\bibliography{reference}{}

\end{document}